\documentclass[twocolumn,prl, aps,superscriptaddress,showpacs]{revtex4-1}

\usepackage{mathptmx}
\usepackage{subfigure}
\usepackage{dcolumn}
\usepackage{amsmath,amssymb}
\usepackage{bm}
\usepackage{color}
\usepackage{latexsym}
\usepackage{epstopdf}
\usepackage{color}
\usepackage[english]{babel}
\usepackage{latexsym}

\usepackage{psfrag,graphicx} 
\usepackage{epsf} 
\usepackage{subfigure} 
\usepackage{amsmath} 
\usepackage{amssymb} 
\usepackage{amsfonts}
\usepackage{bm}
\usepackage{natbib}
\usepackage{epstopdf}
\usepackage{appendix}

\definecolor{mygrey}{gray}{0.35}
\definecolor{myblue}{rgb}{0.2,0.2,0.8}
\definecolor{myzard}{cmyk}{0,0,0.05,0}
\definecolor{mywhite}{rgb}{1,1,1}
\definecolor{myred}{rgb}{1,0.,0.3}

\usepackage[colorlinks=true,citecolor=myblue,linkcolor=myred]{hyperref}

\def\be{\begin{equation}}
\def\ee{\end{equation}}
\def\ba{\begin{align}}
\def\enda{\end{align}}
\def\bi{\begin{itemize}}
\def\ei{\end{itemize}}

 \def\ee{\mathord{\rm e}}

 \def\ee{\mathord{\rm e}}

\renewcommand{\ee}{{\rm e}}

\def\beq{\begin{equation}}
\def\beq{\begin{equation}}
\def\eeq{\end{equation}}

\begin{document}

\title[Short Title]{Quantum Simulation of the Haldane Phase Using Trapped Ions}

\author{I. Cohen}
\author{A. Retzker}
\affiliation{Racah Institute of Physics, The Hebrew University of Jerusalem, Jerusalem 91904, Givat Ram, Israel}

\date{\today}


\begin{abstract}

{A proposal to use trapped ions to simulate spin-one XXZ antiferromagnetic (AFM) chains 
as an experimental tool to explore the Haldane phase is presented. We explain how to reach the Haldane phase adiabatically, demonstrate the 
robustness of the ground states to noise in the magnetic field and Rabi frequencies, and propose a way to detect them using their 
characterizations: an excitation gap and exponentially decaying correlations, a nonvanishing nonlocal string order and a 
double degenerate entanglement spectrum. Scaling up to higher dimensions and more frustrated lattices, we obtain 
richer phase diagrams, and we can reach spin liquid phase, which can be detected by its entanglement entropy which 
obeys the boundary law.}
\end{abstract}

\maketitle


Quantum simulation\cite{QSim,Qsim2} is a promising direction for the exploration of many-body physics. 
Many different experimental platforms have been designed to conduct quantum simulation experiments, 
ranging from cold atomic gases  in optical lattices\cite{Qsim3,Jaksch}, trapped ions\cite{QsimIon}, cavity QED \cite{Hartmann,Cho,Chen}, 
superconducting circuits\cite{Neeley} to linear optics\cite{Alan-photonic simulators}.
The simulation of spin chains has attracted special attention~\cite{Simon et al.,segnstock,Kumar,Ma,Neeley}. In particular trapped ions have been 
targeted as a promising system for the realization of spin chain Hamiltonians\cite{Wunderlich_Sim,Tobias_Sim}. 
Theoretical proposals~\cite{diego,Christof1} have also triggered experimental work
~\cite{Schaetz1,Monroe1,Monroe2,Zippilli,Britton 2D}. 


In the last three decades, a great deal of theoretical effort has been invested in understanding the physics of one dimensional 
one-spin AFM Heisenberg chains; namely, the Haldane phase. According to Haldane's conjecture ~\cite{2 Haldane conjecture} 
integer-spin AFM Heisenberg chains, as described by the following Hamiltonian
 \begin{equation}
H=\sum_{i}S_{x}^{i}S_{x}^{i+1}+S_{y}^{i}S_{y}^{i+1}+\lambda S_{z}^{i}S_{z}^{i+1}+D\left(S_{z}^{i}\right)^{2} 
\label{Haldane Hamiltonian}
  \end{equation}
will have short range correlation functions, and a finite gap between the ground and excited states. 
In contrast, half-integer spin chains will have long-range correlations and will be gapless. 
Kennedy and Tasaki ~\cite{3 Z2*Z2a,4 Z2*Z2a} showed that the Haldane phase is related to the breaking of a hidden
$Z_{2}\times Z_{2}$ symmetry that is introduced by a non-local unitary transformation, 
and followed by a nonzero, nonlocal string order parameter. Recently, Oshikawa {\it et al.}~\cite{5 
symmetry protected topological phase,6 Berg 2} showed that in the most general case, the
Haldane phase can be defined by a symmetry-protected double-degeneracy, even if the string order and 
the gapless edge states are absent. We take advantage of these qualities of Haldane phase to form 
the Hamiltonian adiabatically and to measure its ground states.

In what follows, we propose a scheme to simulate spin one XXZ AFM systems.
To the best of our knowledge, there 
is no experimentally convenient system for exploring the Haldane phase or higher dimensional integer-spin Heisenberg AFM systems. 
Complicated experiments 
using neutron scattering from a one-spin AFM 
Hiesenberg chain CsNiCl$_{3}$ have measured the excitation spectrum and found an energy gap~\cite{8 neutron scattering}. 
However, a trapped ions platform for the quantum simulation of the Haldane phase, enables much more to be said and examined, 
in a much more experimentally feasible way. 

\begin{figure}
   \centering
   \includegraphics[width=1\columnwidth]{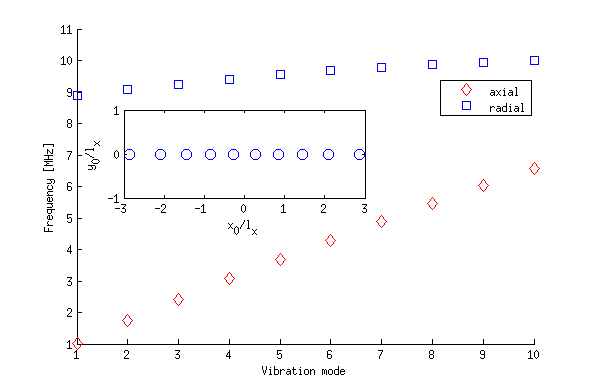}
	\caption {{\bf Vibration modes.} Radial and axial vibration frequencies for N=10 trapped ions in a linear formation
	(legend) calculated when the trap frequencies are $\omega_x=1$MHz $\omega_y=10$MHz. The inset shows the equilibrium 
	positions of the ions.}
  \label{vibration modes}
  \end{figure}

{\em The model ---} In our model we have $N$ ions of mass $m$ and charge $e$, in a linear trap with frequencies 
$\omega_{x},\omega_{y},\omega_{z}$. The MHz trap frequencies are tunable so as to construct the desired geometric shape 
of the ion lattice. A Coulomb interaction causes the trapped ions to vibrate around fixed points 
$r_{i\alpha}=r_{i\alpha}^{0}+\Delta r_{i\alpha}$ of the $i^{th}$ ion in the $\alpha$ direction, whose geometric formation 
is determined by the equilibrium between the trapping forces and the Coulomb repulsion. In the harmonic approximation the 
vibration Hamiltonian ~\cite{9 Trapped ions Hamiltonian,10 S=1/2 AFM Qsim} is
 \begin{equation}
H_{vib}=\sum_{i,\alpha}\left(\frac{1}{2m}p_{i,\alpha}^{2}+\frac{1}{2}m\omega_{\alpha}^{2}\Delta
r_{i,\alpha}^{2}\right)+\frac{e^{2}}{2}\sum_{i,j,\alpha,\beta}V_{ij\alpha\beta}\Delta r_{i\alpha}\Delta r_{j\beta}
  \end{equation} 
  By solving the quadratic problem we obtain the normal modes of the vibration
  (Fig.~\ref{vibration modes}) $ M_{i,n}^{\alpha} $ and $\nu_{n}^{\alpha}$ 
  which are the eigenstates and the eigenvalues of the $n^{th}$ mode and the $i^{th}$ ion in  $\alpha$ direction 
  respectively. Thus, the ion displacements are represented by these normal modes as 
  $\Delta r_{i\alpha}=\sum_{n}M_{i,n}^{\alpha} {\sqrt{2m\nu_{n}^{\alpha}}} {\left (b_{n\alpha}^{\dagger}
  +b_{n\alpha} \right)} $, while the vibration Hamiltonian becomes
  \begin{equation}  H_{vib}=\sum_{n,\alpha}\nu_{n}^{\alpha}b_{n\alpha}^{\dagger}b_{n\alpha}.
  \label{vibration Hamiltonian}
  \end{equation}
  
\begin{figure}
	\centering
	\includegraphics[width=1\columnwidth]{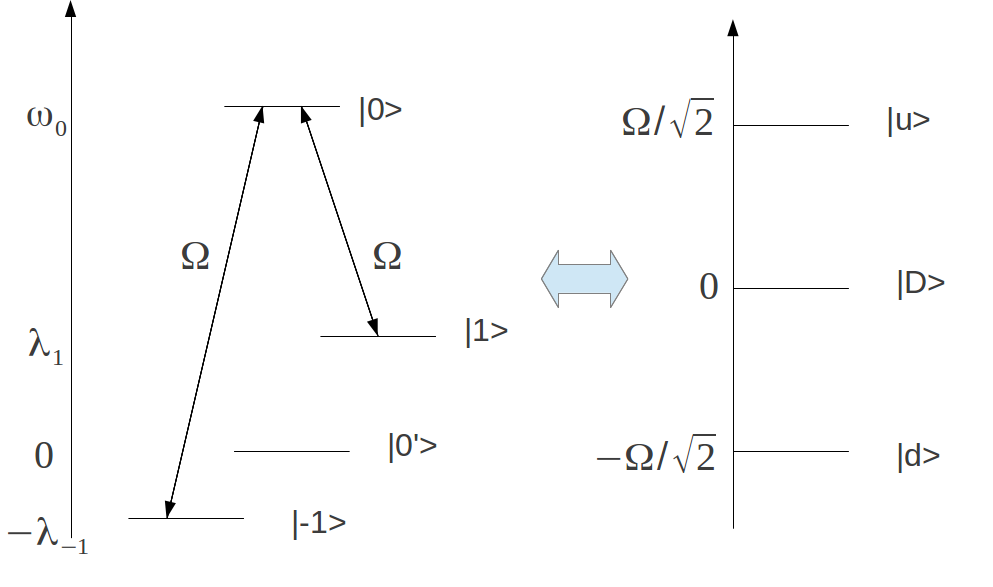}
	\caption{ {\bf Spin Energy levels.} Hyperfine energy levels and the driving fields (left side) that are 
	used for preparing the effective Hamiltonian $H_{XY}$
  in the dressed state basis (right side).} 
 \label{Energy levels}
\end{figure}

 The proposed scheme concentrates on the microwave quantum computing 
 setup~\cite{Christof1,11 2 halves HF ions MW.gate,12 2 halves HF ions MW. gate - Winni} where the spin degrees of freedom are modelled via the 
 hyperfine structure (Fig. \ref{Energy levels} left). The energy levels configuration is produced by the F=0 singlet 
state $\left|0\right\rangle$ and the triplet 
F=1 states $\left|-1\right\rangle$, $\left|0'\right\rangle$ and $\left|1\right\rangle$ according to their spin 
projection on the z direction. Two resonant microwave fields with the same Rabi frequency $\Omega$ drive the transitions 
$\left|\pm1\right\rangle \longleftrightarrow\left|0\right\rangle$. According to their relative phases, the dressed 
states, which will play the role of the simulated S=1 spins, are obtained. 
For the vanishing initial phase difference, the dressed 
state basis is the eigenvector set of $F_{x}$ (the projection of the hyperfine spin on the x axis), and is expanded by 
the original basis $\left\{ \left|1\right\rangle ,\left|0\right\rangle ,\left|-1\right\rangle \right\}$ as the following:
$ \left|u\right\rangle =\frac{1}{\sqrt{2}}\left(\frac{\left|1\right\rangle +\left|-1\right\rangle }{\sqrt{2}}+
  \left|0\right\rangle \right),$ $\left|D\right\rangle =\frac{-\left|1\right\rangle +\left|-1\right\rangle }{\sqrt{2}},$ $\left|d\right\rangle =\frac{1}{\sqrt{2}}\left(\frac{\left|1\right\rangle +\left|-1\right\rangle }{\sqrt{2}}-\left|0
  \right\rangle \right),$ with the eigenvalues $\frac{\Omega}{\sqrt{2}},0,-\frac{\Omega}{\sqrt{2}}$  respectively 
  (Fig. \ref{Energy levels} right).
  
  To induce the spin-spin interaction which is achieved by the creation of a virtual phonon in one ion and its 
  annihilation in another ion, there should be a term in the Hamiltonian that couples the spin and the 
  vibration. Therefore, the very small Lamd-Dicke parameter of microwave sources requires the use of large and stable 
  magnetic field gradients~\cite{Christof1,14 f2} where the field is polarized in the $z$ direction, 
  resulting in adding the term $g\mu_{B}F_{z}^{i}\partial_{\alpha}B_{z}\Delta r_{i,\alpha}$, where $g$ is the Lande 
  g-factor and $\mu_{B}$ is the Bohr magneton.
  In the interaction picture according to the hyperfine structure, the Hamiltonian 
  has the form of 
  \begin{equation}
   \label{H start}
   H_{I}=\sum_{i,n,\alpha}\frac{\Omega}{\sqrt{2}}F_{x}^{i}+\nu_{n}^{\alpha}\left( b^{\dagger}_{n\alpha}b_{n\alpha}+
   \eta_{in}^{\alpha}\left( b^{\dagger}_{n\alpha}+b_{n\alpha}\right)F_{z}^{i}\right)
  \end{equation}
   where $\eta_{in}^{\alpha}=\frac{g\mu_{B}\partial_{\alpha}B_{z}M_{i,n}^{\alpha}}{\sqrt{2m\nu_{n}^{\alpha}}}\ll1$.

 To manifest the  coupling of the phonons to the spins a polaron-like 
  transformation is performed  $U_{p}=e^{-P}$, $P=-\sum_{n,i,\alpha}\frac{\eta_{in}^{\alpha}F_{z}^{i}}{\nu_{n}^{\alpha}}
  \left(b_{n\alpha}^{\dagger}-b_{n\alpha}\right)$. After transforming to the interaction picture according to 
  the carrier transition 
  $\frac{\Omega}{\sqrt{2}}F_{x}$ and the vibration Hamiltonian (Eq.~\ref{vibration Hamiltonian}), 
  the spin-dependent force in the dressed basis is obtained: 
  \begin{equation}
  \label{H_int} 
  H_{I}=\frac{\Omega}{2\sqrt{2}}\sum_{n,i,\alpha}\eta_{in}^{\alpha}
\left(S_{+}^{i}e^{i\frac{\Omega}{\sqrt{2}}t}-S_{-}^{i}e^{-i\frac{\Omega}{\sqrt{2}}t}\right)
\left(b_{n\alpha}^{\dagger}e^{i\nu_{n}^{\alpha}}-b_{n\alpha}e^{-i\nu_{n}^{\alpha}}\right)
 \end{equation}
 
  Expanding the Dyson series of the time propagator to the second order in $H_{I}$
  we obtain two terms in the Hamiltonian 
  $H_{eff}=H_{res}+H_{XY}$:
  \begin{equation}
 \label{XY hamiltonian}
 \centering 
 H_{res}=\sum_{j,n,m,\alpha,\beta}J_{jnm,\alpha\beta}^{res}S_{z}^{j}\left(b_{n\alpha}^{\dagger}b_{m\beta}+
 \frac{1}{2}\delta_{\alpha,\beta}\delta_{n,m}\right)
 e^{-i\left(\nu_{n}^{\alpha}-\nu_{m}^{\beta}\right)t}
   \end{equation}
     \begin{equation}
 H_{XY}=\sum_{i,j}J_{ij}^{eff}\left(\left(S_{x}^{i}S_{x}^{j}+S_{y}^{i}S_{y}^{j}\right)\left(1-\delta_{i,j}\right)-
 \frac{\left(S_{z}^{j}\right)^{2}}{2}\delta_{i,j}\right)\\ 
 \end{equation}
 where
   \begin{equation}
 \label{XY coefficients}
 \centering
J_{jnm}^{res}=\sqrt{2}\Omega\left(\frac{\Omega}{4}\right)^{2}\eta_{jn}^{\alpha}\eta_{jm}^{\beta}
\left(\frac{1}{\left(\frac{\Omega}{\sqrt{2}}\right)^{2}-\left(\nu_{n}^{\alpha}\right)^{2}}+
\frac{1}{\left(\frac{\Omega}{\sqrt{2}}\right)^{2}-\left(\nu_{m}^{\beta}\right)^{2}}\right)
\end{equation}
\begin{equation}
J_{ij}^{eff}=\left(\frac{\Omega}{2}\right)^{2}\sum_{n,\alpha}\eta_{in}^{\alpha}\eta_{jn}^{\alpha}
\frac{2\nu_{n}^{\alpha}}{\left(\frac{\Omega}{\sqrt{2}}\right)^{2}-\left(\nu_{n}^{\alpha}\right)^{2}}
\propto\left|
\vec{r_{i}^{0}}-\vec{r_{j}^{0}}\right|^{-\alpha}_{i\neq j}\\
 \end{equation}
 which gives the power-law spin-spin interactions, instead of the theoretical simplification of nearest neighbors 
 interactions ~\cite{diego}. Although a constant D term, like the last term in Eq. ~\ref{Haldane Hamiltonian}, was obtained, 
 we would like to pursue a tunable D coefficient. As we will see later in this article our model can be 
 decoupled from $H_{res}$ since its ground states belong to the decoherence-free subspace.
  
  \begin{figure}
	\centering
	\includegraphics[width=1\columnwidth]{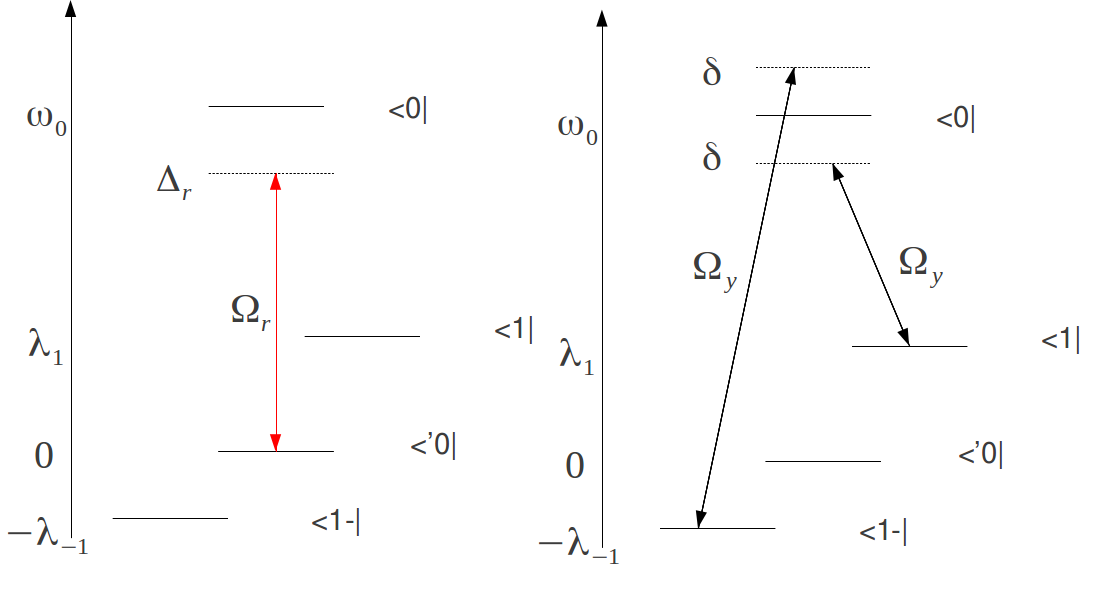}
	\caption{ {\bf Driving fields for creating the D and $\lambda$ coefficients.} Left side: Generating the D term in 
	Eq.~\ref{Haldane Hamiltonian} is done by an A.C. Stark shift via the red detuned 
	$\Delta_{r}$ transition. Right side: The transitions that are required for generating $\Omega'S_{y}\sin\theta$ 
	and applying the ``trick`` where $\delta=\left(\frac{\Omega}{\sqrt{2}}-\Omega'\cos\theta\right)$ is the detuning, 
	$ \pm\frac{\pi}{2} $ are the initial phases and $\Omega_{y}=\sqrt{2}\Omega'\sin\theta$ is the Rabi frequency.}
 \label{D coefficient}
\end{figure}

{\it Generation of the anisotropy D term ---} Generating a tunable D term is achieved by applying 
an A.C. Stark shift using an additional microwave field corresponding to the transition between the states $\left|0'\right\rangle
\longleftrightarrow\left|0\right\rangle$ with a red detuning $\Delta_{r}$ (Fig.~\ref{D coefficient} left).
In the dressed-state basis $\left|0\right\rangle$ is transformed to $\frac{1}{\sqrt{2}}\left(\left|u\right\rangle-\left|d\right\rangle\right)$, 
therefore by moving to the interaction picture according to the carrier transition $\frac{\Omega}{\sqrt{2}}S_{z}$ as was mentioned above, 
which is followed by an expansion of Dyson series of the time propagator to the second order,
all the off-diagonal terms are suppressed using the rotating wave approximation (RWA), if we assume that
$\frac{\Omega_{r}}{2\sqrt{2}}\ll\frac{\Omega}{\sqrt{2}}\ll\Delta_{r}$. As a consequence, the anisotropy 
D'=$\frac{\Omega_{r}^{2}}{8\Delta_{r}}$ term is obtained, which is slightly different from D since the model is $\theta$ rotated
while generating $\lambda$ term.

\begin{figure}
	\centering
	\includegraphics[width=1\columnwidth]{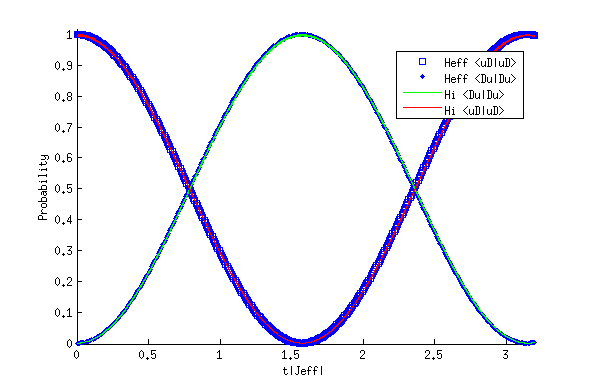}
	\caption{ {\bf Dynamics of two trapped ions for an initial state $\left|uD\right\rangle $ } 
	The probability of finding the system in states $\left|uD\right\rangle $ and $\left|Du\right\rangle $ 
	calculated by both 
	$H_{eff} $ and $H_{Int} $, using the following physical values: $\eta=0.03$, $\nu=4$MHz, $\Omega=6\sqrt{2}\nu$, 
	$\theta=1.47\sim\frac{\pi}{2}$, $\Omega'=0.6\nu$ (Rabi frequency of $S_{z,\theta}$ for the “trick”), $D'=-18.5$KHz 
	(the coefficient of $S_{z}^{2}$ obtained by the A.C Stark shift). We obtain the Haldane Hamiltonian with: $Jeff=1.85$KHz
	D coefficient $D=4.35$ and $\lambda$ coefficient $\lambda=0.989$. 
 	This confirms that Eq.\ref{Heff_final} was obtained appropriately.}
 \label{Hi vs Heff}
\end{figure}

  {\it Generation of the Ising-like $\lambda$ term ---} Instead of producing the Ising-like anisotropy term $\lambda S_{z}^{i}S_{z}^{i+1}$
  explicitly as was done with the anisotropy D term, 
  we will use the following ``trick``. We will add $H_{2}=\Omega'S_{z,\theta}=\Omega'
  \left(S_{z}\cos\theta-S_{y}\sin\theta\right)$ to the Hamiltonian, which is $\Omega'F_{x,\theta}=\Omega'\left(F_{x}\cos\theta-F_{y}
  \sin\theta\right)$ in the original basis. This is done by generating each term separately. The first term 
  $\Omega'F_{x}\cos\theta=\Omega'S_{z}\cos\theta$ can be set aside from the previous microwave transitions that 
  produce $H_{XY}$. The second term $\Omega'S_{y}\sin\theta$ (Fig.~\ref{D coefficient} right) is 
  produced by applying two microwave 
  fields corresponding to the transitions $\left|\mp1\right\rangle \longleftrightarrow\left|0\right\rangle$ with 
  $\pm\left(\frac{\Omega}{\sqrt{2}}-\Omega'\cos\theta\right)$ detunings, $\pm\frac{\pi}{2}$ initial phases and the same 
  Rabi frequency $\sqrt{2}\Omega'\sin\theta$ respectively, after transforming to the interaction picture according to 
  $\left(\frac{\Omega}{\sqrt{2}}-\Omega'\cos\theta\right)F_{x}$, and neglecting the fast rotating terms. The Rabi 
  frequency should be much smaller than the original Rabi frequency namely $\Omega'\ll\Omega$, so the undesired spin to 
  vibration coupling caused by the polaron-like transformation will be negligible. By applying a $\theta$
  rotation around the $x$ axis, the operators are trasformed as follows: $S_{z,\theta}\rightarrow S_{z}$,
  $S_{z}\rightarrow S_{z}\cos\theta+S_{y}\sin\theta$, $S_{y}\rightarrow S_{y}\cos\theta-S_{z}\sin\theta$ and $S_{x}$
  is not changed, and the new dressed-state basis is $\theta$ rotated around the x axis as well. If we move to the interaction 
  picture corresponding to the new term we have built  and use RWA where we want $\frac{\eta_{in}^{\alpha}\Omega}{2\sqrt{2}}\ll\Omega' $ 
  we will end up with the following effective Hamiltonian
   \begin{equation}
   \begin{split} 
   \label{Heff_final}
 H_{eff}=\sum_{i\neq j}{J_{ij}^{eff}}\left\{ \left(S_{x}^{i}S_{x}^{j}+S_{y}^{i}S_{y}^{j}\right)\left(\frac
     {1+\cos^{2}\theta}{2}\right)+S_{z}^{i}S_{z}^{j}\frac{\sin^{2}\theta}{2}\right\}\\
     +\sum_{i}\left(\frac{\Omega_{r}^2}{8\Delta_{r}}-
     \frac{J_{ii}^{eff}}{2}\right)
    \left(\frac{1+\cos^{2}\theta}{2}-\sin^{2}\theta\right)  \left(S_{z}^{i}\right)^{2}
  \end{split} 
    \end{equation}
  which is a one-spin XXZ AFM Hamiltonian with power-law spin-spin interactions. In Fig.~\ref{Hi vs Heff} 
  we illustrate the 
  comparison between the Hamiltonian of the spin dependent force  $H_{Int}$ (Eq.~\ref{H_int}, to which were added 
  all the previous steps (Fig.~\ref{D coefficient})) and the effective Hamiltonian $ H_{eff} $ for two trapped ions, 
  using physically reasonable values. 

  \begin{figure}
	\centering
	\includegraphics[width=1\columnwidth]{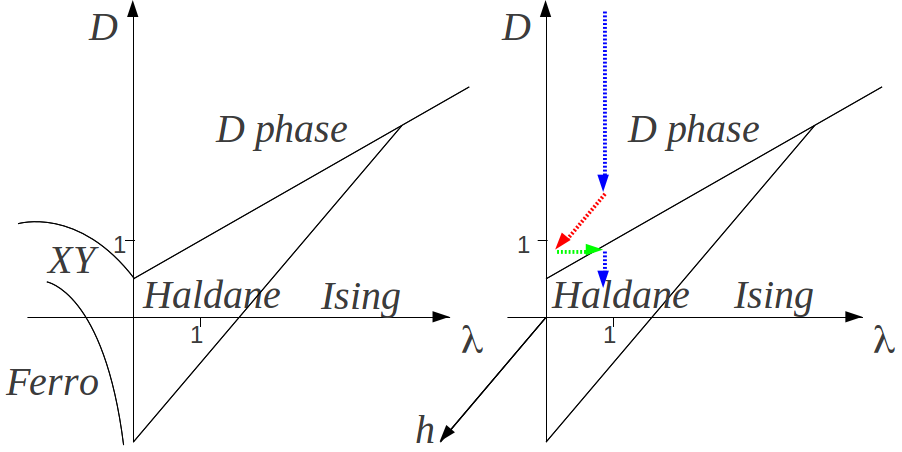}
	\caption{{\bf Phase diagram and adiabatical path} Phase diagram of the Haldane phase, where the phases are the large 
	D phase, the Haldane phase, the Ising-like phase, the XY phase and the ferromagnetic phase. The colored arrows 
	represent the adiabatical path from the large D phase to Haldane phase.The blue arrow stands for lowering D coefficient, 
	the red arrow stands for swiching on the symmetry breaking h term while still lowering D and the green arrow stands for 
	swiching off the h term while lowering D.}
 \label{phase diagram}
\end{figure}
  
  {\it Reaching the Haldane phase adiabatically ---} In order to prepare the system in the Haldane phase 
  and to reach its ground state, we first start with the large $D$ Hamiltonian and generate its ground topologically 
  trivial state which is a tensor product of local $\left|D\right\rangle$ states in every site ~\cite{12 2 halves HF ions 
  MW. gate - Winni,Christof1}. Then we switch on the required Hamiltonian adiabatically 
  ({\it i.e.} slower than the energy gap) by 
  lowering the $D$ coefficient. For an infinite chain, a problem arises when we reach a second order phase 
  transition in the phase diagram (Fig.~\ref{phase diagram}) where the energy gap closes \cite{Sachdev} and the adiabatic approximation cannot hold. 
  To overcome this   obstacle we take advantage of the fact that the Haldane phase is asymmetry-protected topological phase. 
  Far enough from the 
  $D\rightarrow H$ phase transition we will adiabatically turn on a perturbation in the Hamiltonian that breaks all the 
  symmetries in this system. Then, when we are in the region of the Haldane phase we will adiabatically turn this perturbation 
  off. In that way we will not cross any second order phase border, and hence we will always stay in the adiabatic 
  approximation. 
  
The Haldane phase is protected by the following symmetries: a bond centered spatial inversion $\vec{S}_{j}\rightarrow\vec
{S}_{-j+1}$, a time reversal symmetry $\vec{S}_{j}\rightarrow-\vec{S}_{j}$ or the dihedral $D_{2}$ symmetry which is the 
$\pi$ rotations around x, y and z axes. In order to break all these symmetries we add a perturbation term 
$H_{pert}=-h\sum_{i}\left(-1\right)^{i}S_{z}^{i}$. This term can be produced by individual 
addressing~\cite{14 
f2} using a microwave frequency comb with a staggered phase. Taking advantage of the magnetic field gradient along the 
chain axis, each ion experiences a different Zeeman splitting and only the right frequency from the comb can interact with 
it. 

Before crossing the phase transition, the symmetry breaking perturbation should be turned on adiabatically, while still 
lowering the D coefficient in the Hamiltonian. Then, the perturbation should be turned off adiabatically, reaching the plane 
h=0 in the Haldane phase domain. If the time duration of this procedure is shorter than the coherence time, 
the ground state in the Haldane phase should be achieved with high fidelity.

  {\it Detecting and measuring the ground states ---} As a 
  topological phase, the Haldane phase does not obey the Landau paradigm and cannot be characterized by a local order parameter. However 
  there are other properties that can characterize it: $1)$ an excitation gap and exponentially decaying correlations 
  $C_{ij}^{\alpha}=\left\langle S_{i}^{\alpha}S_{j}^{\alpha}\right\rangle -\left\langle S_{i}^{\alpha}\right
  \rangle^{2}$, $2)$ a nonvanishing nonlocal string order $O_{string}^{\alpha}\left(H\right)=\lim_{|i-j|\rightarrow
  \infty}\left\langle -S_{i}^{\alpha}\exp\left[i\pi\sum_{l=i+1}^{j-1}S_{l}^{\alpha}\right]S_{j}^{\alpha}\right\rangle$, 
  where $\left\langle \right\rangle$ denotes the expectation value in the ground state, $3)$ a double degenerate entanglement 
  spectrum, obtained by dividing the systems into two parts, tracing out one of them and diagonalizing the reduced density 
  matrix~\cite{entanglement spectrum}.

Using trapped ions as a platform for quantum simulations allows us measure 
every spin state with high fidelity and accuracy. The tomography of an exponentially growing Hilbert space with the size of 
the system is time-consuming. Yet, efficient tailored reconstruction methods ~\cite{17 f4} make it posible to calculate 
the correlation functions and the string order. Thanks to the dipole power-law interaction in our model, we should find a power-law tail in 
the correlation functions in addition to the exponentially decay $C_{ij}^{\alpha}=Ae^{-\frac{\left|i-j
\right|}{\xi}}+B\left|i-j\right|^{-a}$. By implementing a less time consuming method, we can calculate the entanglement spectrum of the 
ground state and determine whether it is double degenerate, according to the Haldane phase signature.
  
  {\it Robustness of the ground states to noise ---} Preparing the system in the Haldane phase is done 
  adiabatically. Note that this can take more time than the time scale set by the noise sources. The main noise sources here are 
  the fluctuating magnetic field in the $z$ direction and the fluctuations in the Rabi frequencies of the driving fields. If 
  during the path we represented above, the ground states are in the decoherence free subspace, the specific subspace of 
  Hilbert space that is invariant under these fluctuations (to the first order), we will be able to walk on that path 
  adiabatically. 

The dressed basis used for representing the spins is the eigenstates of $F_{x}$ $\left\{ \left|u\right\rangle ,\left|D
\right\rangle ,\left|d\right\rangle \right\}$. These states are robust to the magnetic noise in the $z$ direction, as 
$\left\langle s\right|o_{\left(t\right)}F_{z}\left|s\right\rangle =0$ for $\left|s\right\rangle \in\left\{ \left|u
\right\rangle ,\left|D\right\rangle ,\left|d\right\rangle \right\}$. Moreover, the ground state of the large D phase which is 
the topologically trivial state of $\left|D\right\rangle$ states in every site and the ground state of the Haldane phase~\cite{
3 Z2*Z2a,4 Z2*Z2a} which has the same number of sites occupied with $\left|u\right\rangle$ and $\left|d\right
\rangle$ are robust to the fluctuations in the Rabi frequencies, as $o_{\left(t\right)}F_{x}\left|GS\right\rangle
=o_{\left(t\right)}S_{z}\left|GS\right\rangle=0$.
The remarkable structure of the ground states explains why our model is decoupled from 
$H_{res}\propto S_z^i $ (Eq.~\ref{XY hamiltonian}), and we do not have to work 
hard to  dynamically decouple it 
by a $\pi $-pulse sequence~\cite{dynamical decouple}.
  
  {\it Higher Dimensions ---} We can also scale up to higher dimensions with more frustrated geometries of the quantum spin liquid 
  phase where the interaction excitations are strong. Using the linear Paul trap, by lowering the transverse trap 
  frequency $\omega_{z}$ enough, we observe a phase transition from the linear formation to a frustrated zigzag n-ladder 
  formation ~\cite{18 d=2 lattice}. As the fabrication technology progresses, arbitrary geometries could be produced for the ions lattice. 
The ground state of the spin liquid phase is characterized (and as a result 
  can be detected) by the entanglement entropy that obeys the boundary law $S=a\sigma_{R}-\gamma$, where $\sigma_{R}$ is 
  the $d-1$ dimensional volume surrounding the region, and $\gamma=\log d$ is a fixed value in the topologically ordered 
  phase, which is independent of the lattice's geometry ~\cite{19 f5,20 f6,21 f7}. 

{\em Summary ---} We proposed a quantum simulation scheme for one-spin Heisenberg AFM systems. 
The Hamiltonian can be generated adiabatically, starting from the large D phase until it reaches the spin-liquid phase 
(or the Haldane phase in $d=1$). The ground states are robust to magnetic noise, decoupled from the fluctuations in the Rabi 
frequencies and can be detected by their characterization as mentioned above.

{\em Acknowledgements ---} We thank E. Berg and D. Orgad for useful discussions and acknowledge the support of the European Commission (STREP EQuaM).


\newpage
\end{document}